\documentclass
[twocolumn,aps,prd,amsmath,showpacs,amssymb,floatfix]
{revtex4-1}
\usepackage{CJK}                     
\usepackage[dvips]{graphicx}
\usepackage{bm}                      
\usepackage{mathptmx}                
\usepackage{dcolumn}                 
\usepackage{xcolor}                  %

\begin{document}

\def\bea{\begin{eqnarray}} \def\eea{\end{eqnarray}}
\def\be{\begin{equation}} \def\ee{\end{equation}}
\def\bal#1\eal{\begin{align}#1\end{align}}
\def\bse#1\ese{\begin{subequations}#1\end{subequations}}
\def\rra{\right\rangle} \def\lla{\left\langle}
\def\eps{\epsilon}
\def\ms{M_\odot}
\def\mmax{M_\text{max}}
\def\fm3{\;\text{fm}^{-3}}
\def\pt{p_\text{th}}
\def\et{\eps_\text{th}}
\def\gt{\Gamma_\text{th}}



\title{
Hot neutron stars with microscopic equations of state
}

\begin{CJK*}{GB}{gbsn} 

\author{Jia-Jing Lu (陆家靖)$^{1,2}$}        
\author{Zeng-Hua Li (李增花)$^1$} \email[]{zhli09@fudan.edu.cn}
\author{G. F. Burgio$^2$}
\author{H.-J. Schulze$^2$}

\affiliation{$^1$%
Institute of Modern Physics,
Key Laboratory of Nuclear Physics and Ion-beam Application (MOE),
Fudan University, Shanghai 200433, P.R.~China\\
$^2$INFN Sezione di Catania, Dipartimento di Fisica,
Universit\'a di Catania, Via Santa Sofia 64, 95123 Catania, Italy}

\date{\today}

\begin{abstract}
We study the properties of hot beta-stable nuclear matter
using equations of state
derived within the Brueckner-Hartree-Fock approach
at finite temperature
including consistent three-body forces.
Simple and accurate parametrizations of the finite-temperature
equations of state are provided.
The properties of hot neutron stars are then investigated
within this framework,
in particular the temperature dependence of the maximum mass.
We find very small temperature effects and analyze the interplay
of the different contributions.
\end{abstract}


\maketitle
\end{CJK*}

\section{Introduction}

The recent first observation of a neutron star (NS) merger event,
GW170817 \cite{merger},
has opened new possibilities to understand the properties of the extremely
hot and dense environment that is created by the fusion of two NSs,
and represents a transitory state to either collapse to a black hole or
the formation of a very heavy NS \cite{revrez,paste}.

The theoretical modeling of this system requires first of all
the knowledge of the nuclear equation of state (EOS)
under the extreme conditions
of several times normal nuclear matter density
$\rho_0 \approx 0.17\fm3$,
and temperatures of tens of MeV.
It is the motivation of this work to provide realistic microscopically founded
EOSs for this purpose.
We will therefore extend to finite temperature several microscopic EOSs
that have been derived within the Brueckner-Hartree-Fock (BHF) formalism
\cite{bhf,book99,bb12}
based on realistic two-nucleon and three-nucleon forces
\cite{bbb,zuotbf,litbf}.
They feature reasonable properties at (sub)nuclear densities
in agreement with nuclear-structure phenomenology \cite{litbf,cpltbf,pheno},
and are also fully compatible with recent constraints obtained from the
analysis of the GW170817 event \cite{asylam}.

Apart from the application to simulations of merger events,
such finite-temperature EOSs are also
relevant for the modeling of compact stellar objects
like protoneutron stars
\cite{burrows,prakash,ponsevo,ourpns,bsaa}
and supernovae \cite{super}.

A particularly important feature of any NS EOS is the associated
maximum stable NS mass.
Currently a lower limit for the cold EOS is due to
the observation of NSs with above two solar masses \cite{mmax},
recently updated to $\mmax>2.17\pm0.1 \ms$ \cite{mmax2},
whereas the analysis of GW170817
(in particular its delayed decay to a black hole)
has permitted to establish approximate upper limits of about $2.2\,\ms$
\cite{mmmax}
for the maximum mass of a static cold NS.
Two important physical effects influence the estimate of the maximum mass
of a cold static NS from the properties of the transient object
(hypermassive NS)
created by the merger:
First, the remnant is (differentially) rotating fast,
which allows temporarily a higher metastable mass
(of about 20 percent)
than for the nonrotating object
\cite{hashi,goussard,schaab,nsrot}.
Second, the remnant is hot,
and therefore its maximum mass is different from the one of the cold object.
For this estimate the temperature dependence of the EOS becomes essential.

There is therefore a possible mass range of the hot rotating
transitory metastable object that depends on the finite-temperature EOS.
Of course the precise determination of this feature requires sophisticated
simulations \cite{simu,baus,pascha,lalit}
and we will in this article only give very simple estimates of the effect,
using and comparing our different EOSs.

A few finite-temperature nuclear EOSs for astrophysical simulations
are now available \cite{ourpns,bsaa,ls,shen,hempel,dd,fsu,nl3,sfho},
and the predictions for the effects of temperature on stellar stability
are conflicting:
Relativistic-mean-field (RMF) models usually predict increasing stability
(maximum mass) with temperature \cite{baum,prakash,kaplan},
whereas BHF results \cite{ourpns,bsaa}
indicate in general a slight reduction of the maximum mass.
We will try to analyze in some detail this feature.

Our paper is organized as follows.
In Sec.~\ref{s:bhft} we discuss the finite-temperature BHF approach
and the fits of our finite-temperature results for the free energy.
The composition of stellar matter and the EOS are presented in Sec.~\ref{s:eos},
along with the equations of stellar structure.
The numerical results regarding temperature effects
on EOS and maximum mass
are then illustrated in Sec.~\ref{s:res},
and conclusions are drawn in Sec.~\ref{s:end}.

\section{Formalism}
\label{s:bhf}

\subsection{Brueckner-Bethe-Goldstone theory at finite temperature}
\label{s:bhft}

\begin{figure*}[t]
\vspace{-7mm}\hspace{-1mm}
\centerline{\includegraphics[scale=0.55]{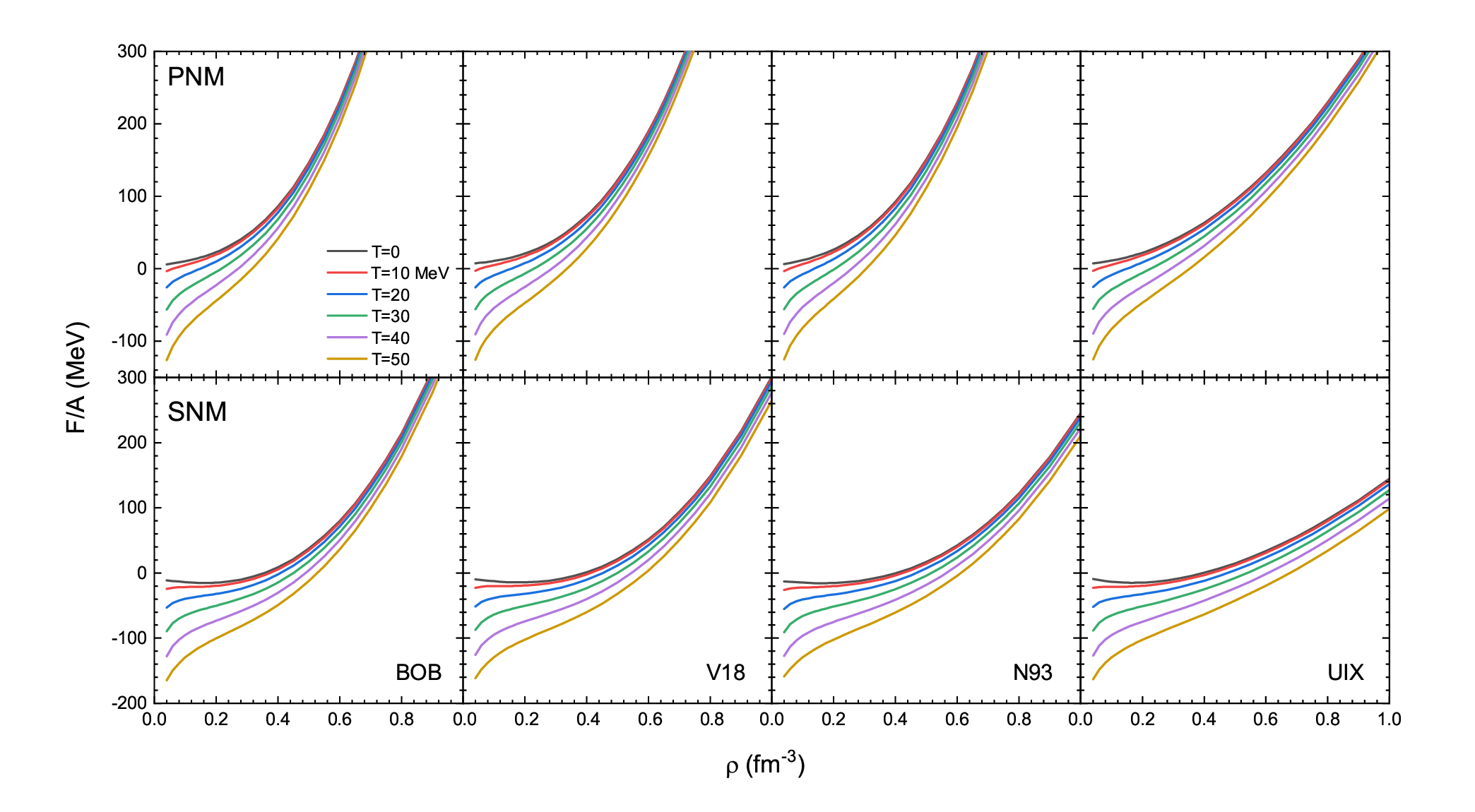}}
\vspace{-11mm}
\caption{
Free energy per nucleon as a function of nucleon density
for symmetric (lower panels) and pure neutron (upper panels) matter
with different EOSs,
for temperatures ranging from 0 to 50 MeV in steps of 10 MeV.
}
\label{f:bf}
\end{figure*}

The free energy density of hot nuclear matter consists of two contributions,
\be
 f = f_N + f_L \:,
\label{e:f}
\ee
where $f_N$ is the nucleonic part
and $f_L$ denotes the contribution of leptons $e,\mu,\nu_e,\nu_\mu$,
and their antiparticles.
In the present work,
we employ the BHF approach for asymmetric nuclear matter at finite temperature
\cite{bloch,book99,lej86,baldo,ourpns,bsaa}
to calculate the nucleonic contribution.
The essential ingredient of this approach is the interaction matrix $K$,
which satisfies the self-consistent equations
\be
  K(\rho,x_p;E) = V + V \;\text{Re} \sum_{1,2}
 \frac{|12 \rangle (1-n_1)(1-n_2) \langle 1 2|}
 {E - e_1-e_2 +i0} K(\rho,x_p;E) \:
\label{eq:BG}
\ee
and
\be
 U_1(\rho,x_p) = {\rm Re} \sum_2 n_2
 \langle 1 2| K(\rho,x_p;e_1+e_2) | 1 2 \rangle_a \:,
\label{eq:uk}
\ee
where $x_p=\rho_p/\rho$ is the proton fraction, and
$\rho_p$ and $\rho$ are the proton and the total baryon density, respectively.
$E$ is the starting energy and
$e(k) \equiv k^2\!/2m + U(k)$ is the single-particle (s.p.) energy.
The multi-indices 1,2 denote in general momentum, isospin, and spin.

Several choices for the realistic nucleon-nucleon ($NN$) interaction $V$
are adopted in the present calculations \cite{litbf}:
the Argonne $V_{18}$ \cite{v18},
the Bonn~B (BOB) \cite{bonn},
and the Nijmegen~93 (N93) \cite{nij},
and compatible three-nucleon forces as input.
More precisely,
the BOB and N93 are supplemented with microscopic TBF
employing the same meson-exchange parameters as the two-body potentials
\cite{grange,zuotbf,litbf},
whereas $V_{18}$ is combined either with a microscopic or a phenomenological TBF,
the latter consisting of an attractive term due to two-pion exchange
with excitation of an intermediate $\Delta$ resonance,
and a repulsive phenomenological central term \cite{uix,bbb}.
They are labeled as V18 and UIX, respectively,
throughout the paper and in all figures.
The TBF are reduced to an effective two-body force
and added to the bare potential in the BHF calculation,
see Refs.~\cite{grange,zuotbf,litbf} for details.

At finite temperature, $n(k)$ in Eqs.~(\ref{eq:BG}) and (\ref{eq:uk})
is a Fermi distribution.
For a given density and temperature, these equations have
to be solved self-consistently along with the following equations for
the auxiliary chemical potentials $\tilde{\mu}_{n,p}$,
\be
 \rho_i = 2\sum_k n_i(k) =
 2\sum_k {\left[\exp{\Big(\frac{e_i(k)-\tilde{\mu}_i}{T}\Big)}
 + 1 \right]}^{-1} \:.
\label{eq:ro}
\ee

To save computational time and simplify the numerical procedure,
in the following we employ the so-called
frozen-correlations approximation \cite{ourpns,baldo}, i.e.,
the correlations at $T \neq 0$
are assumed to be essentially the same as at $T = 0$.
This means that the s.p.~potential $U_i(k)$ for the component $i$
at finite temperature is approximated by the one calculated at $T=0$.
Within this approximation,
the nucleonic free energy density has the following simplified expression,
\be
 f_N = \sum_{i=n,p} \left[ 2\sum_k n_i(k)
 \left( {k^2\over 2m_i} + {1\over 2}U_i(k) \right) - Ts_i \right] \:,
\label{eq:f}
\ee
where
\be
 s_i = - 2\sum_k \Big( n_i(k) \ln n_i(k) + [1-n_i(k)] \ln [1-n_i(k)] \Big)
\label{eq:entr}
\ee
is the entropy density for the component $i$ treated as a free Fermi gas with
spectrum $e_i(k)$.
It turns out that the assumed independence is valid to a good accuracy
\cite{ourpns,baldo},
at least for not too high temperature, $T\lesssim30\;\text{MeV}$.

We stress that the BHF approximation,
both at zero and finite temperature,
does not fulfill the Hugenholtz-Van Hove theorem \cite{baldo},
and therefore the following procedure has to be adopted
in order to derive all necessary thermodynamical quantities
in a consistent way from the total free energy density $f$,
namely
one defines the ``true" chemical potentials $\mu_i$,
pressure $p$, and internal energy density $\eps$ as
\bea
 \mu_i &=& \frac{\partial f}{\partial \rho_i} \:,
\\
 p &=& \rho^2 {\partial{(f/\rho)}\over \partial{\rho}}
 = \sum_i \mu_i \rho_i - f \:,
\label{e:eosp}
\\
 \eps &=& f + Ts \:,\quad
 s = -{{\partial f}\over{\partial T}} \:.
\label{e:eose}
\eea

For illustration, we display in Fig.~\ref{f:bf} the
nucleonic free energy per nucleon,
$F/A=f_N/\rho$,
as a function of the baryon density $\rho$,
obtained following the above discussed procedure,
for symmetric nuclear matter ($x_p=1/2$, SNM)
and pure neutron matter ($x_p=0$, PNM),
and the different EOSs we are using,
for several values of temperature between 0 and 50 MeV.
At $T=0$ the free energy coincides with the internal energy
and the corresponding SNM curve is just the usual
nuclear matter saturation curve.
The temperature effect is less pronounced for PNM due to the larger
Fermi energy of the neutrons at given density.
We notice that the results are ordered with increasing stiffness of the EOS
as UIX, N93, V18, BOB.

For practical use,
we provide analytical fits of the free energy $F/A(\rho, T)$ for SNM and PNM.
We find that in both cases
the following functional forms provide excellent parametrizations
of the numerical results in the required ranges of density
($0.05\fm3 \lesssim \rho \lesssim 1\fm3$)
and temperature ($5\;\text{MeV} \leq T \leq 50\;\text{MeV}$):
\bea
{F\over A}(\rho,T) &=&
 a \rho + b \rho^c + d
\nonumber\\&&
 +\, \tilde{a} t^2 \rho
 + \tilde{b} t^2 \ln(\rho)
 + ( \tilde{c} t^2 + \tilde{d} t^{\tilde{e}} )/\rho  \:,
\label{e:fitf}
\eea
where $t=T/(100\;\text{MeV})$ and $F/A$ and $\rho$ are given in
MeV and $\fm3$, respectively.
The parameters of the fits are listed in Table~\ref{t:fit}
for the different EOSs we are using.
The rms deviations of fits and data are better than 1 MeV for all EOSs.

For the asymmetric matter case,
it turns out that the dependence on proton fraction
can be very well approximated by a parabolic law,
as at zero temperature \cite{bombaci,bsaa}:
\bea
 {F\over A}(\rho,T, x_p) &\approx&
 {F\over A}(\rho,T, 0.5)
\\&&\nonumber
 + (1-2x_p)^2 \Bigg[ {F\over A}(\rho,T, 0) - {F\over A}(\rho,T, 0.5) \Bigg]
\:.
\label{e:parab}
\eea
Therefore, for the treatment of the beta-stable case,
it is only necessary to provide parametrizations for SNM and PNM.
For convenience we provide in the supplemental material \cite{suppl}
the complete EOS tables in the parameter space of
temperature, baryon density, and proton fraction.

\begin{table}
\caption{
Parameters of the fit for the free energy per nucleon $F/A$,
Eq.~(\ref{e:fitf}),
for symmetric nuclear matter (SNM) and pure neutron matter (PNM)
and the different EOSs used.}
\medskip
\def\myc#1{\multicolumn{1}{c}{$#1$}}
\renewcommand{\arraystretch}{1.2}
\begin{ruledtabular}
\begin{tabular}{lr|rrrr|rrrrr}
     &     & \myc{a} & \myc{b} & \myc{c} & \multicolumn{1}{c|}{$d$} &
     $\tilde{a}$ & $\tilde{b}$ & $\tilde{c}$ & $\tilde{d}$ & $\tilde{e}$ \\
\hline
 BOB& SNM & -65 & 498 & 2.67 & -9 & -124 & 203 & -105 & 122 & 2.20 \\
 BOB& PNM &  57 & 856 & 2.91 &  4 &  -85 & 152 &  -32 &  43 & 2.47 \\
\hline
 V18& SNM & -60 & 369 & 2.66 & -8 & -147 & 209 &  -66 &  85 & 2.32 \\
 V18& PNM &  37 & 667 & 2.78 &  6 &  -91 & 154 &  -52 &  62 & 2.28 \\
\hline
 N93& SNM & -42 & 298 & 2.61 &-12 & -142 & 211 &  -64 &  87 & 2.35 \\
 N93& PNM &  67 & 743 & 2.71 &  4 &  -95 & 154 &  -35 &  46 & 2.44 \\
\hline
 UIX& SNM &-174 & 323 & 1.61 & -4 & -186 & 199 & -136 & 153 & 2.16 \\
 UIX& PNM &  24 & 326 & 2.09 &  6 & -117 & 153 &  -85 &  94 & 2.16 \\
\end{tabular}
\end{ruledtabular}
\label{t:fit}
\end{table}

\subsection{Composition and EOS of hot stellar matter}
\label{s:eos}

The purpose of this article is to evaluate the effect of the intrinsic
temperature dependence of the nuclear EOS caused by the strong interaction.
However, finite temperature also affects the composition of stellar matter
governed by the weak interaction.
In this article we study exclusively beta-stable
and neutrino-free nuclear matter,
assuming that the temporal evolution is slow enough to justify these assumptions.
This might not necessarily be a good approximation for
merger simulations \cite{alf}.

In beta-stable nuclear matter
the chemical potential of any particle $i=n,p,l$ is uniquely determined
by the conserved quantities baryon number $B_i$, electric charge $Q_i$,
and weak charges (lepton numbers) $L^{(e)}_i$, $L^{(\mu)}_i$:
\be
 \mu_i = B_i\mu_n
 + L^{(e)}_i\mu_{\nu_e}  + L^{(\mu)}_i\mu_{\nu_\mu} \:.
\label{mufre:eps}
\ee
For stellar matter containing nucleons and leptons as relevant degrees
of freedom,
the chemical equilibrium conditions read explicitly
\be
 \mu_n - \mu_p = \mu_e = \mu_\mu \:.
\label{beta:eps}
\ee
At given baryon density $\rho$,
these equations have to be solved together with the
charge-neutrality condition
\be
 \sum_i Q_i \rho_i = 0 \:.
\label{neutral:eps}
\ee
The various chemical potentials are obtained from the total
free energy density $f$, Eq.~(\ref{e:f}),
\bea
 \mu_i(\{\rho_j\}) &=&
 \left. \frac{\partial f}{\partial \rho_i} \right|_{\rho_{j\neq i}} \:.
\label{mun:eps}
\eea
Using the hadronic and leptonic chemical potentials,
one can calculate the composition of beta-stable stellar matter,
and then the total pressure $p$ and the internal energy density $\eps$,
through the usual thermodynamical relations
expressed by Eqs.~(\ref{e:eosp},\ref{e:eose}).
Once the EOS $p(\eps)$ is specified,
the stable configurations of a NS can be obtained from the
well-known hydrostatic equilibrium equations
of Tolman, Oppenheimer, and Volkov \cite{shapiro}
for pressure $p(r)$,
enclosed gravitational mass $m(r)$,
and baryonic mass $m_B(r)$
\bea
 {dp\over dr} &=& -\frac{Gm\eps}{r^2}
 \frac{\big( 1+ p/\eps \big) \big( 1 + 4\pi r^3p/m \big)}
 {1-2Gm/r} \:,
\\
 \frac{dm}{dr} &=& 4\pi r^2\eps \:,
\\
 \frac{dm_B}{dr} &=& 4\pi r^2 \frac{\rho m_N}{\sqrt{1-2Gm/r}}  \:,
\label{e:tov}
\eea
where $m_N=1.67\times 10^{-24}$g is the nucleon mass
and $G= 6.67408 \times 10^{-8}\text{cm}^3\text{g}^{-1}\text{s}^{-2}$
the gravitational constant.
For a chosen central value of the energy density,
the numerical integration of these equations provides the
mass ($M,M_B$) -- radius ($R$) relations.

The solution of these equations depends obviously on the
temperature profile $T(r)$.
In any realistic simulation of an astrophysical scenario at finite temperature
(supernova, protoneutron star, merger),
the TOV equations are therefore embedded in a detailed
and self-consistent dynamical simulation of the temperature evolution.
We cannot perform such detailed studies here,
but our current aim is just to identify the global effect of
finite temperature on the stability of a NS merger remnant,
as motivated in the Introduction.
This will serve as a preparation for a better qualitative understanding of
future detailed simulations based on our finite-temperature EOSs.

We therefore assume simply a constant temperature inside the star and
attach for the outer part a cold crust
given in Ref.~\cite{nv} for the medium-density regime
($0.001\fm3<\rho<0.08\fm3$),
and in Refs.~\cite{bps,fmt}
for the outer crust ($\rho<0.001\fm3$).
The maximum-mass domain that we are interested in,
is hardly affected by the structure
of this low-density transition region \cite{ourpns,bsaa}.

\section{Results}
\label{s:res}

\begin{figure}[t]
\vspace{-9mm}
\centerline{\includegraphics[scale=0.38]{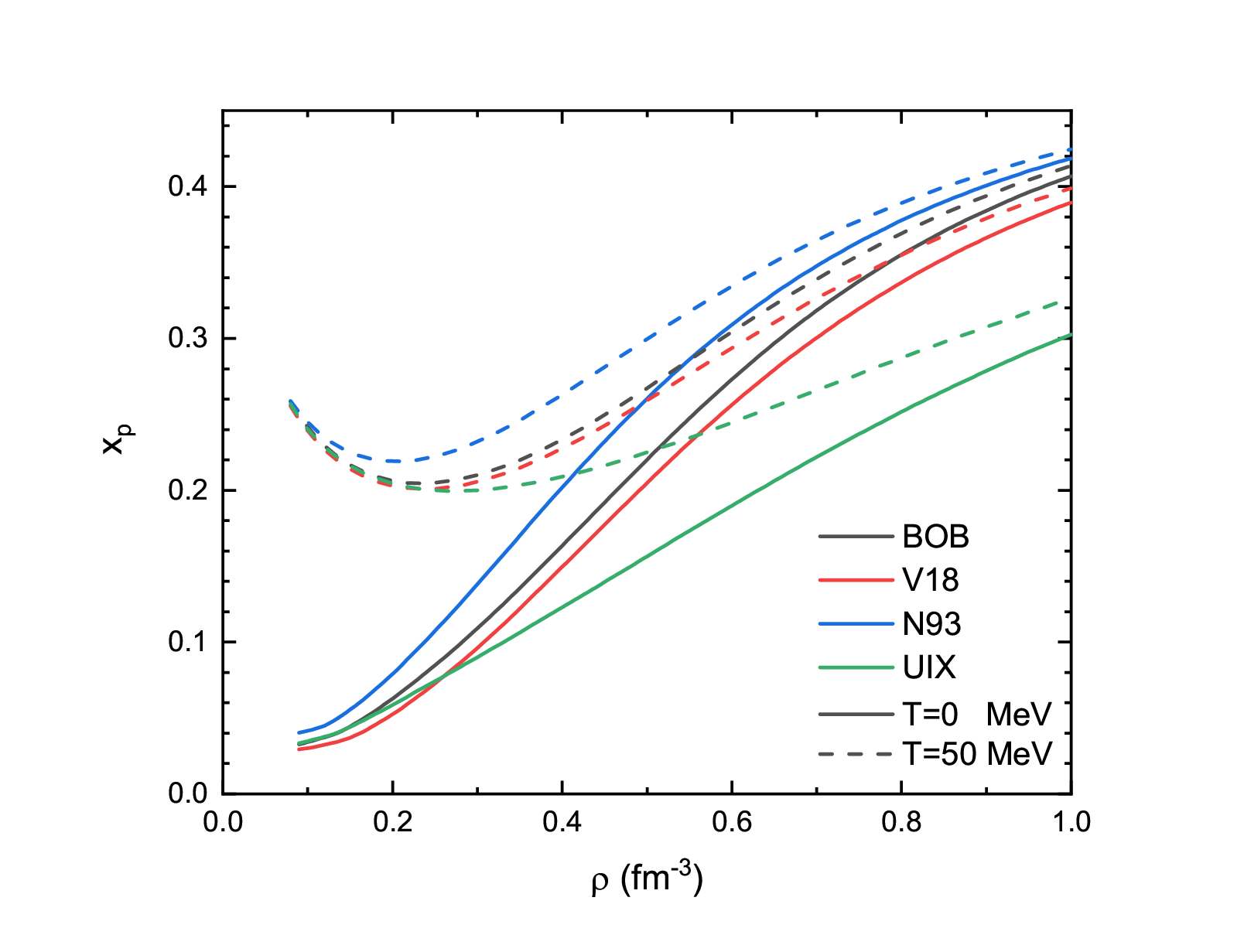}}
\vspace{-7mm}
\caption{
Proton fractions in beta-stable matter
at the temperatures $T=0$ (solid curves) and 50 MeV (dashed curves)
for the different EOSs.}
\label{f:xt}
\end{figure}

\begin{figure}[t]
\vspace{-6mm}
\includegraphics[scale=0.32]{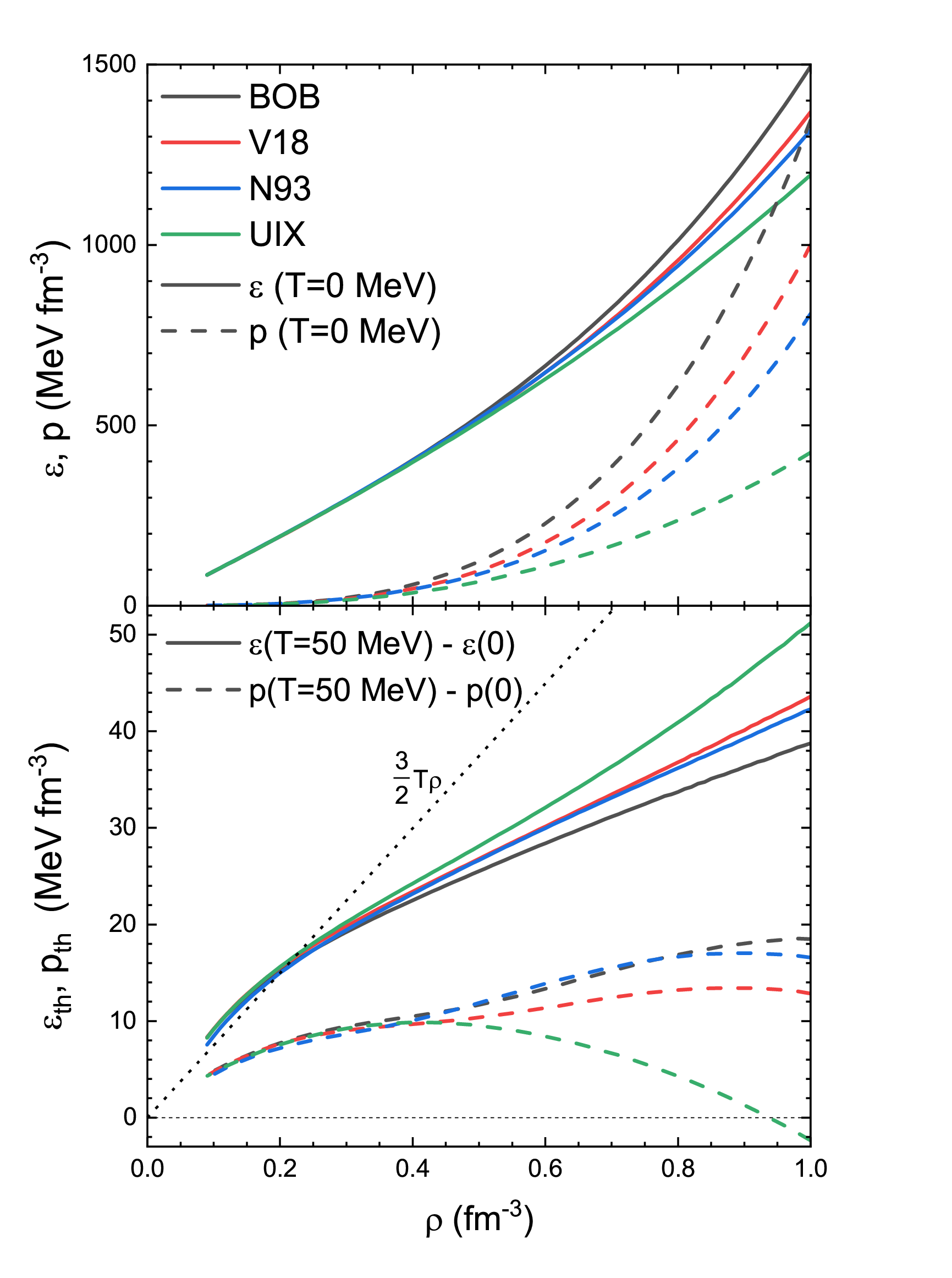}
\vspace{-10mm}
\caption{
Internal energy density $\eps$ (solid curves)
and pressure $p$ (dashed curves)
of beta-stable matter
at $T=0$ (upper panel)
and changes of those quantities at $T=50$ MeV
for the different EOSs (lower panel).}
\label{f:p}
\end{figure}

\begin{figure}[t]
\vspace{-3mm}
\centerline{\includegraphics[scale=0.23]{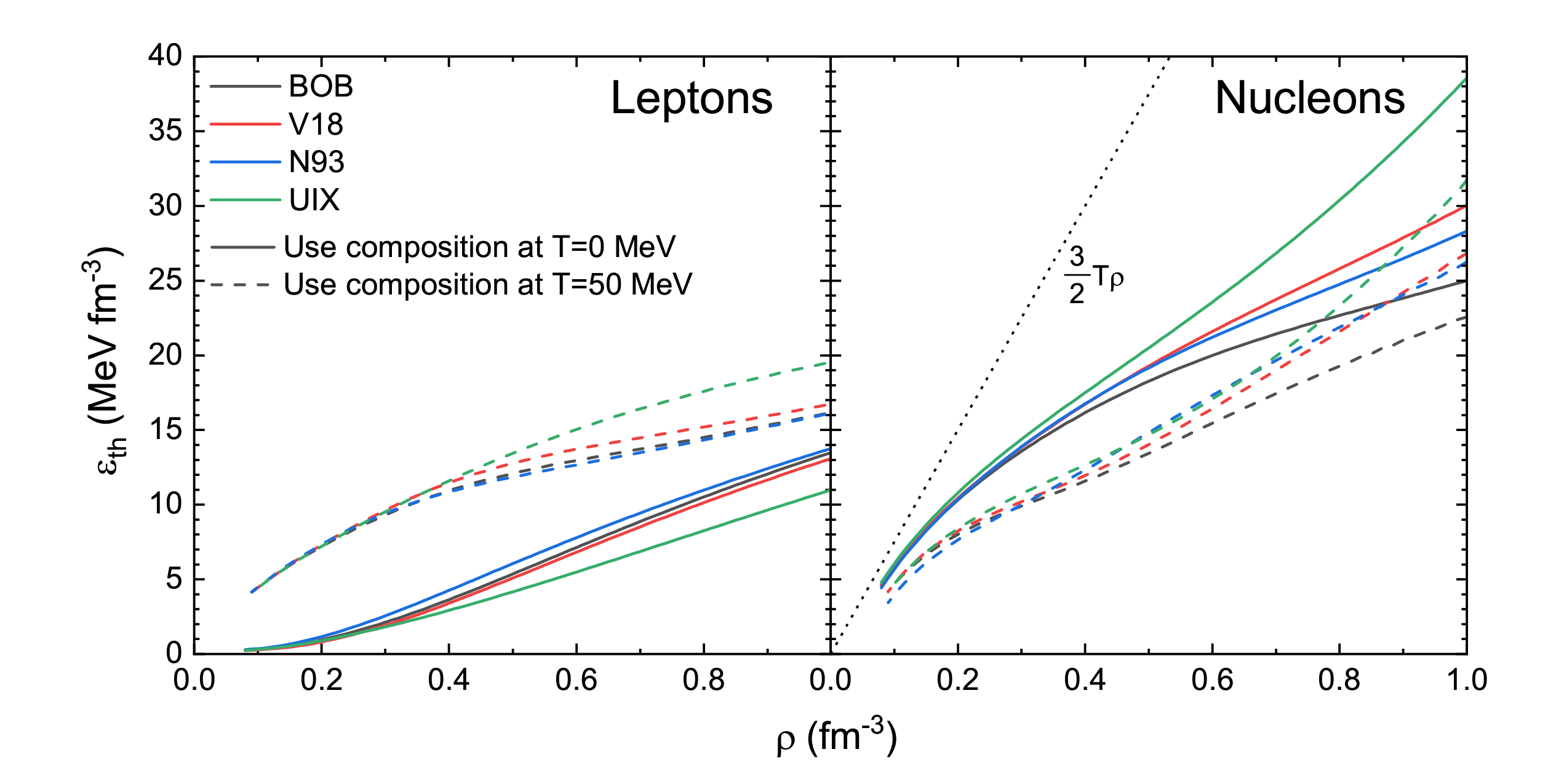}}
\vspace{-4mm}
\caption{
Lepton and nucleon contributions to the
thermal internal energy density of beta-stable matter
at $T=50\,$ MeV for the different EOSs.
The solid (dashed) curves employ the particle fractions of cold (hot) matter,
see Fig.~\ref{f:xt}.
}
\label{f:eps}
\end{figure}

\begin{figure}[t]
\vspace{-8mm}\hspace{3mm}
\centerline{\includegraphics[scale=0.38]{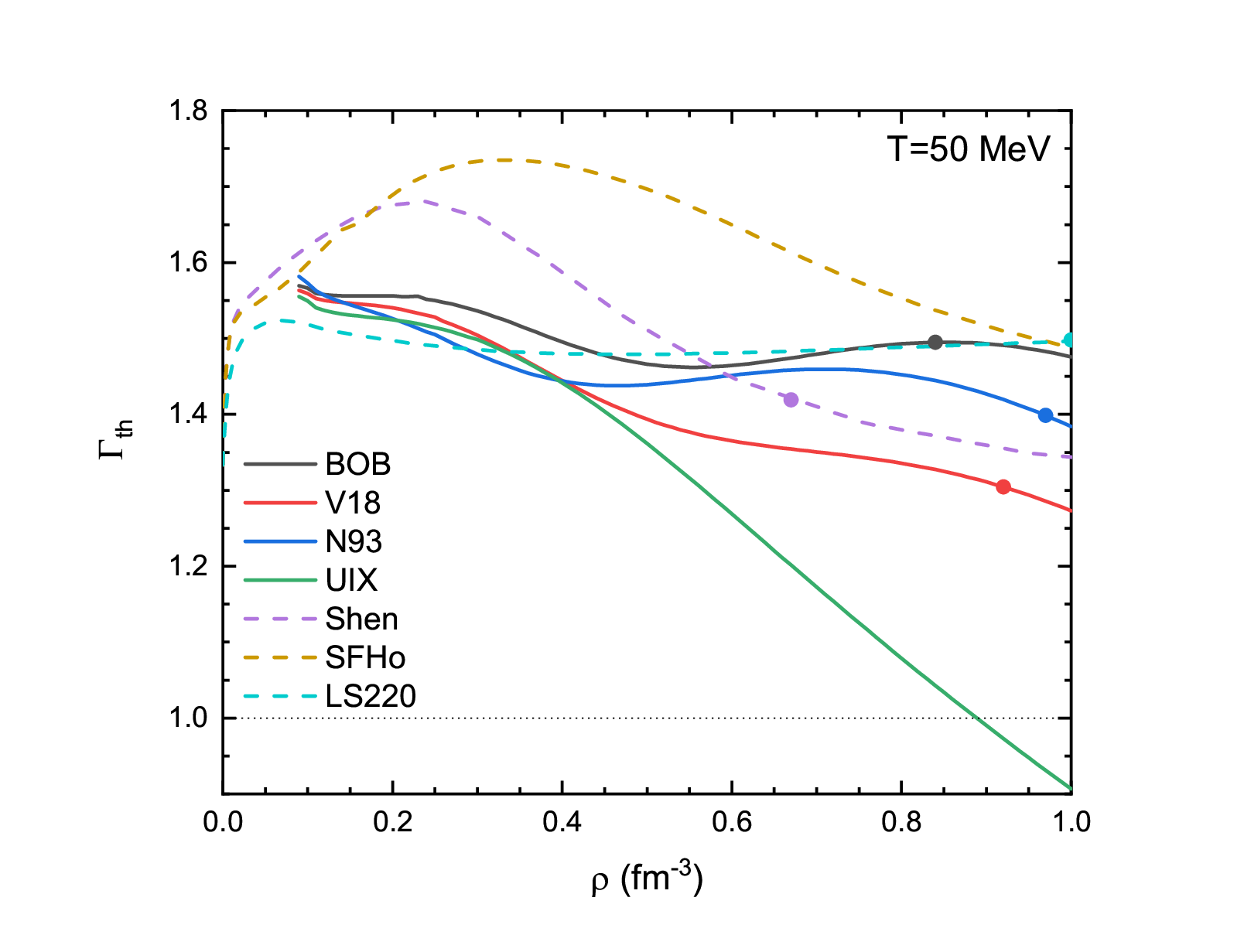}}
\vspace{-10mm}
\caption{
Adiabatic index at $T=50$ MeV
obtained for different EOSs.
The markers indicate the central densities of the maximum-mass stars.
For this comparison (anti)muons have been disregarded in the stellar composition.
}
\label{f:gam}
\end{figure}

\begin{figure*}[t]
\vspace{-3mm}\hspace{-3mm}
\centerline{\includegraphics[scale=0.275]{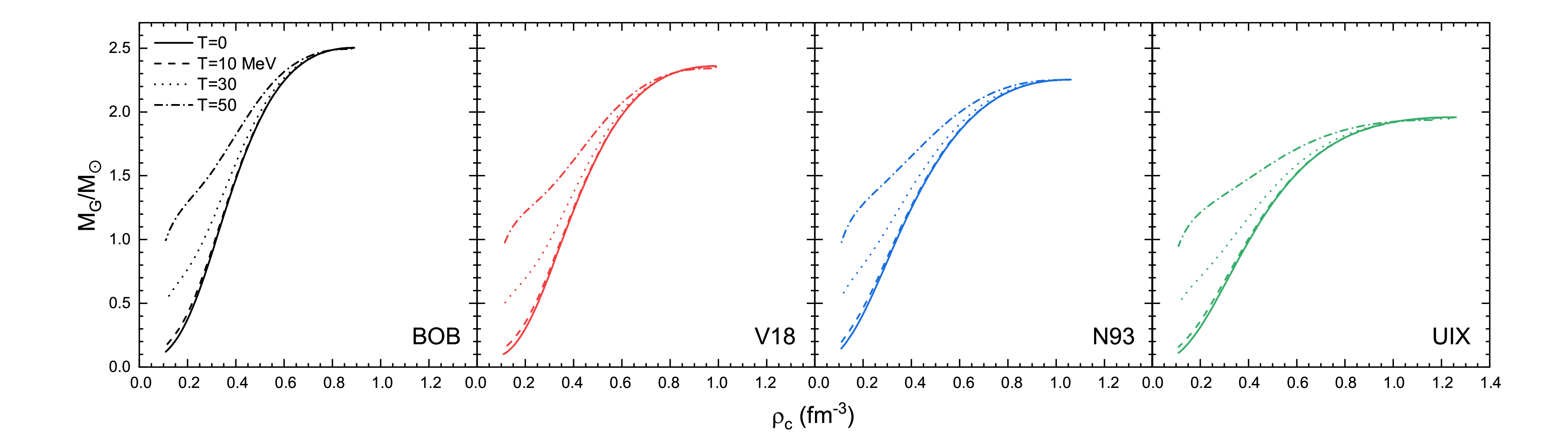}}
\vspace{-5mm}
\caption{
Neutron star gravitational mass vs. central density relations
at $T=0, 10, 30, 50$ MeV, for the different EOSs.}
\label{f:m}
\end{figure*}

In the following we present the results of our numerical calculations
regarding the composition of hot NS matter and the structure of NSs.

\subsection{Composition of stellar matter}

A main characteristic of stellar nuclear matter is its proton fraction,
which is displayed in Fig.~\ref{f:xt}
as a function of the baryon density for the temperatures $T=0$ and 50 MeV,
obtained with the different EOSs.
We stress that both electrons and muons are taken into account for these results.

First we notice that the different EOSs predict somewhat
different proton fractions at high density,
but all of them exceed the threshold value $x_\text{DU}\approx0.13$
for the opening of the direct Urca cooling reactions in cold matter.
The onset density is comprised in the range 0.3--0.4$\fm3$,
and therefore medium-mass NS can cool down very rapidly,
as illustrated in our recent work \cite{ourcool}.
The predicted NS cooling properties with these EOSs
are compatible with all current cooling data.

As far as the finite-temperature effects are concerned,
we notice that the proton fraction is mainly affected in the low-density region,
where leptons become rather numerous  
as a result of Fermi distributions at finite temperature.
Because of the charge-neutrality condition,
this increases the proton fraction
and thus the isospin symmetry of nuclear matter,
and this counteracts the stiffening of the EOS
due to the individual thermal pressures of the nucleons.
On the other hand,
the increase of the lepton densities with temperature
augments the thermal lepton pressure,
which in turn acts against the effect of increasing isospin symmetry.
We will analyze the interplay between these effects in the following.

\subsection{Pressure and energy density}

Fig.~\ref{f:p} shows the EOS of beta-stable matter $p(\rho)$ and $\eps(\rho)$
obtained with the different EOSs at $T=0$ in the upper panel
and the changes at $T=50$ MeV in the lower panel,
i.e., the thermal pressure
$\pt(T,\rho) \equiv p(T,\rho)-p(0,\rho)$ (dashed curves)
and internal energy density
$\et(T,\rho) \equiv \eps(T,\rho)-\eps(0,\rho)$ (solid curves).
One can clearly see the nonmonotonic density behavior of the thermal pressure
due to three competing effects:
The individual thermal pressures of protons and neutrons
at fixed partial densities are increasing,
but the isospin asymmetry is decreasing with temperature
(see Fig.~\ref{f:xt}),
which reduces the total baryonic pressure.
On the other hand, the increased lepton densities augment the lepton thermal
pressure.

In our approach, the overall thermal effects are small,
of the order of a few percent at high density,
even at the fairly high temperature $T=50$ MeV considered here.
In fact
a simple nonrelativistic ideal-gas approximation for nucleons \cite{bomb}
\be
 \et = \frac{3}{2} T \rho
\ee
(dotted curve in the lower panel of Fig.~\ref{f:p}),
significantly overestimates the thermal effects.
This result is independent of the frozen-correlations approximation
adopted here,
as was demonstrated in Ref.~\cite{bsaa}.

In order to understand in some more detail the previous results,
we show in Fig.~\ref{f:eps} separately the lepton and nucleon
contributions to the thermal energy density at $T=50$ MeV,
obtained in two different ways:
The solid curves show the results obtained with the proton fractions
of cold matter (solid curves in Fig.~\ref{f:xt}),
whereas the dashed curves employ the consistent proton fractions at $T=50$ MeV
(dashed curves in Fig.~\ref{f:xt}).
One observes clearly the competition between the increased lepton contribution
due to the larger lepton=proton fractions at finite temperature,
and the decrease of the nucleonic contribution
due to the larger isospin symmetry.
At low density the former effect is dominant,
but at high density there is strong compensation between both.
The overall result is a slight increase of the total thermal energy density
in beta-stable matter due to the action of the weak interaction
via increased lepton and proton fractions.
This change is small compared to the dominant cause of temperature dependence
by the strong interaction as investigated in Sec.~\ref{s:bhft}.

\subsection{Adiabatic index}

An important quantity often used in NS merger simulations
\cite{janka,baus,simu,bomb,raithel}
is the adiabatic index $\gt$ appearing in the
ideal-fluid approximation 
\be
 \pt(T,\rho) = (\gt-1) \et(T,\rho)
\ee
with a constant $\gt$
(originally chosen as $\gt\approx1.5$ \cite{janka}).
We note that due to the thermal effects analyzed before,
this relation might be strongly violated in our microscopic approach,
in particular for EOSs with relatively small proton fraction (UIX),
where the thermal pressure might even become negative at high density.
In fact, three-dimensional relativistic hydrodynamical calculations
of NS mergers \cite{baus}
have questioned the validity of this approximation in the postmerger phase,
where thermal effects are most relevant.
Strong variations were found in both the oscillation frequency
of the forming hypermassive object,
and the delay time between merging and black hole formation,
with respect to the simulations with a fully consistent treatment of temperature.

To illustrate this issue,
we show in Fig.~\ref{f:gam} the adiabatic index
$\gt=1+\pt/\et$
at $T=50\;$MeV
derived from our results for the different EOSs.
There is clearly an important density dependence
(the temperature dependence turns out to be much less pronounced)
and the average remains even below 1.5,
in particular at high density.
For comparison we also display results of some frequently used RMF models,
namely the LS220 EOS \cite{ls},
the Shen EOS \cite{shen},
and the recent SFHo \cite{sfho}.
We plan to study this problem
more extensively in future detailed merger simulations.

\subsection{Stellar structure}

\begin{figure}[t]
\vspace{-4mm}
\centerline{\hspace{0mm}\includegraphics[scale=0.30,clip]{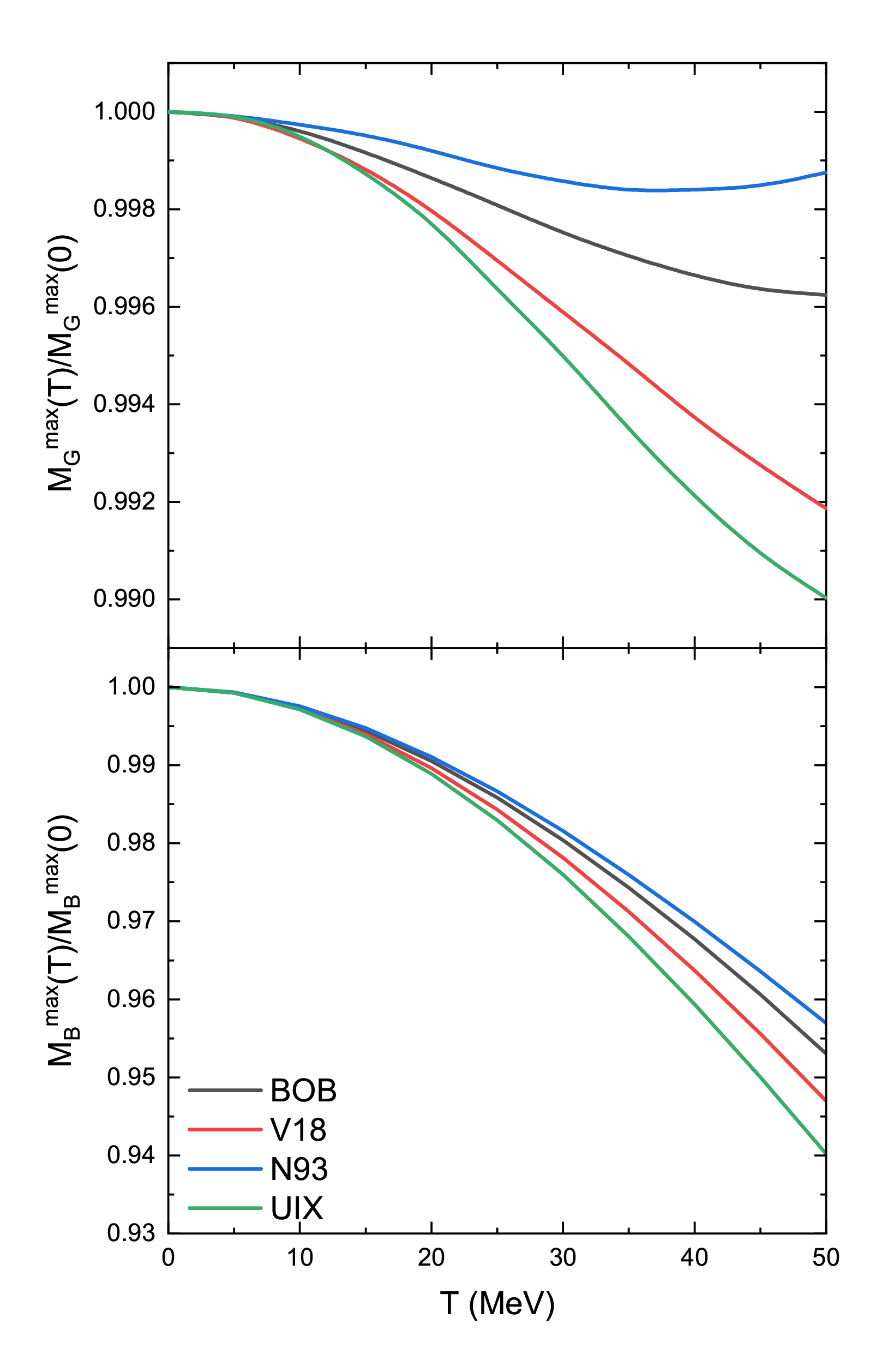}}
\vspace{-7mm}
\caption{
Temperature dependence of the neutron star
gravitational (upper panel) and
baryonic (lower panel)
maximum mass
for the different EOSs.}
\label{f:dm}
\end{figure}

The features of the temperature-dependent EOSs
are reflected in Fig.~\ref{f:m},
where the corresponding gravitational mass vs.~central density relations
are plotted for isothermal stars at $T=0, 10, 30, 50$ MeV.
One notes that the theoretical predictions for the maximum masses depend most
importantly on the nuclear EOS
($\mmax/\ms = $2.50, 2.36, 2.25, 1.96 for BOB, V18, N93, UIX respectively),
whereas the dependence on temperature is nearly negligible,
and summarized in the upper panel of Fig.~\ref{f:dm},
showing the relative change of the maximum mass with temperature.
There is a slight decrease,
reaching about one percent at $T=50$ MeV
for the UIX EOS and less for the other, stiffer EOSs.

However, those results regard the maximum reachable gravitational mass
at finite temperature,
regardless of the (approximate) conservation of baryon number in the
dynamical evolution.
It is therefore also of interest to analyze the behavior of the maximum
baryonic mass $M_B(T)$ with temperature,
and this is shown in the lower panel of Fig.~\ref{f:dm}.
In fact a stronger decrease (up to about 6\%) is observed here,
as less baryons can be bound with increasing temperature.
The much weaker decrease of $M_G(T)$
relative to this behavior is due to the
thermal increase of the internal energy density $\et$,
which adds gravitational mass to the star.

This qualitative analysis is in agreement with the similar one performed
in Ref.~\cite{kaplan} for a number of RMF models at finite temperature,
including the three models featured in Fig.~\ref{f:gam}.
However, in that case
and in the older potential models framework of Ref.~\cite{prakash}
increasing maximum gravitational masses with temperature were reported.
This could be caused by interactions which are mostly local
with a non-local correction, as in Ref.~\cite{prakash},
and therefore thermal effects are included only in the kinetic energy.
In our calculations thermal effects are contained in the whole interaction
part through the single-particle potentials;
thus a completely different temperature dependence may arise.
In fact the relevant adiabatic index is fairly small in our approach,
see Fig.~\ref{f:gam},
while larger values in other models
might be able to cause an increase of the maximum mass.
This will be the subject of further study.

Finally, we remind that the effect of neutrino trapping was completely
disregarded in this schematic investigation,
which focused on the temperature effects of the strong interaction.
The neutrino contributions to thermal energy density and pressure
might also cause a substantial increase of the maximum masses
\cite{ourpns,kaplan,pascha,lalit}.
But for a consistent analysis much more detailed simulations are required.

\medskip\medskip
\section{Summary}
\label{s:end}

In conclusion, we presented microscopic calculations
and convenient parametrizations of the equation of state
of hot asymmetric nuclear matter within the framework of the
Brueckner-Hartree-Fock approach at finite temperature
with different potentials and compatible nuclear three-body forces.
We notice that our results have been obtained in the framework of the
frozen-correlations approximation scheme,
but a more complete calculation is expected to give a similar behavior.

We then investigated the EOS of hot NS matter,
in particular the density dependence of the adiabatic index,
and determined the dependence of the maximum NS mass on temperature
for beta-stable and neutrino-free configurations.
At variance with other available EOSs at finite temperature,
widely used in neutron star merger simulations,
we found a very small maximum mass decrease up to rather large temperatures,
which can be related to the competition between increasing thermal pressures
and increasing isospin symmetry of the stellar nuclear matter.
This small effect would practically justify to disregard
the temperature dependence of the nuclear EOS in merger simulations,
as far as stellar stability is concerned.
To verify this supposition,
we plan to employ the various microscopic EOSs in detailed simulations
to be confronted with future observations of merger events.
This will allow to constrain even more the possible EOS.

\section*{Acknowledgments}

This work is sponsored by
the National Natural Science Foundation of China under Grant
Nos.~11075037,11475045
and the China Scholarship Council, File No.~201806100066.
We further acknowledge partial support from ``PHAROS,'' COST Action CA16214.


\end{document}